\documentclass[%
 reprint,
 amsmath,amssymb, aps, prl,,
]{revtex4-2}

\usepackage{morefloats}
\usepackage{color}
\usepackage{epsfig,graphicx,amsfonts,amsbsy}
\usepackage{amsmath,amsfonts,amsthm,amssymb}
\usepackage{appendix}
\usepackage{bbm}
\usepackage{makeidx}
\usepackage{url}
\usepackage{verbatim}
\usepackage{mathrsfs}

\usepackage{morefloats}
\usepackage{appendix}
\usepackage{bbm}
\usepackage{makeidx}
\usepackage{url}
\usepackage{verbatim}
\usepackage[bookmarksnumbered,pdfpagelabels=true,plainpages=false,colorlinks=true,linkcolor=blue,citecolor=blue,urlcolor=blue]{hyperref}
\usepackage{mathrsfs} 
\usepackage{array}
\usepackage{booktabs}
\usepackage{multirow}
\usepackage{bbm}
\usepackage{tabularx}
\usepackage{cancel,soul}
\usepackage{nicefrac, xfrac}
\usepackage{ulem}
\usepackage{bbold}
\usepackage[version=4]{mhchem}

\newcommand{\bS}{\boldsymbol{{\rm S}}}

\def \fcfm {Departamento de F\'isica, CEDENNA, FCFM, Universidad de Chile, Santiago, 8370448, Chile.}
\def \usach {Departamento de F\'isica, CEDENNA, Universidad de Santiago de Chile, Santiago, 9170124, Chile.}
\def \iai {Instituto de Alta Investigaci\'on, Universidad de Tarapac\'a, Casilla 7D, Arica, Chile}
\def \uay {Departamento de Ciencias Naturales y Tecnología, Universidad de Aysén
Eusebio Lillo 667, Coyhaique 5951616, Chile}

\begin{document}
\title{Controlling the Vacuum: Tunable Magnonic Casimir Effect in Magnetic Moiré Systems}
\author{Alejandro Roldan$^{1}$}
\author{Dante Navarrete$^2$}
\author{Mario A. Castro$^2$}
\author{Sebastián Allende$^{4}$}
\author{Roberto E. Troncoso$^3$}
\author{Alvaro S. Nunez$^{2}$} 
\affiliation{$^1$\uay, $^2$\fcfm, $^3$\iai, $^4$\usach }
\date{\today}

\begin{abstract}
We prove that the quantum fluctuations of the magnetic degrees of freedom of moiré structures can be tuned by adjusting the moiré lattice. Using the Bistritzer-MacDonald approach, we explore the consequences of this fact for the magnonic vacuum energy and the magnonic Casimir force, which become functions of the moiré pattern parameters. This allows a prototypical engine to be run with the vacuum fluctuations as a working agent and, in that way, to furnish a novel type of magneto-mechanical effect. Our calculations focus on the moiré Ladder as an example but can be generalized to any moiré structure in a straightforward fashion. We anticipate that our results belong to a generic property of moiré structures, regardless of the physical nature of the excitations at hand.
\end{abstract}

\maketitle

{\it Introduction.--} Fluctuation-induced forces represent a widespread and inherent result of both quantum and thermal field perturbations occurring within restricted spatial domains \cite{Kardar1999, Dalvit2011}. The concept originated with Casimir's seminal prediction \cite{Casimir1948, Mostepanenko1997, Bordag2009}, which described an attractive force between two ideally conducting plates due to the influence of vacuum electromagnetic fluctuations. This groundbreaking idea has since been extended to a diverse array of physical systems. Among them we can list confined superfluids \cite{Macioek2007, Biswas2010}, Bose–Einstein condensates \cite{Todoshchenko2024}, ferrofluids \cite{Zhang2024}, quantum Hall systems \cite{Muniz2021, Tse2012}, and metamaterials \cite{Rosa2008, Rosa2009}, all of which exhibit similar phenomena. Each of these systems illustrates the emergence of forces from the intrinsic nature of field fluctuations within restricted settings, thereby expanding our comprehension of fluctuation-induced interaction forces. In environments characterized by magnetic order, the low-energy excitations known as magnons form a bosonic field \cite{Yuan2022, Ochoa2019, RoldanMolina2017}. This field exhibits a spectrum and mode structure that is profoundly influenced by boundary conditions, geometrical constraints, and the texture of the magnetic medium. Such sensitivity inherently leads to the phenomenon of the magnonic Casimir effect \cite{RoldanMolina2015, Nakata2023, Nakata2024, Tapia2024}. In this effect, alterations in the magnonic vacuum result in the emergence of forces, torques, and energy shifts between boundaries, defects, or embedded magnetic configurations. This intricate interplay between the magnonic field and its environment highlights the complex nature of quantum interactions at play, offering deeper insights into how fundamental forces manifest in constrained quantum systems.

In recent years, significant progress in the field of magnonics has greatly broadened our capabilities to design, manipulate, and study spin-wave modes at the nanoscale level \cite{Demokritov2012, Pirro2021, Han2024, Faurie2021}. The development of engineered magnetic heterostructures, the creation of patterned waveguides, and the use of reconfigurable textures, such as domain walls, vortices, and skyrmions, have established groundbreaking platforms that enable the precise manipulation of magnon spectra. Within these advanced systems, the complex interaction between geometric confinement and intrinsic interactions, including exchange interactions, dipolar coupling, and the Dzyaloshinskii–Moriya interaction, can substantially alter the density of magnon states. This modification creates optimal conditions for the emergence of fluctuation-mediated forces and torques \cite{Ai2023}, providing new insights and opportunities for exploration in nanoscale magnetic systems. Theoretical predictions indicate that the magnonic Casimir effect could manifest in a variety of ways beyond the conventional static forces. This phenomenon might also lead to shifts in spectral characteristics, exert texture-dependent torques, or induce intricate alterations in the dynamics of collective modes. 

Simultaneously, the pursuit of low-power, wave-based information processing technologies has sparked significant interest in the utilization of magnonic forces as potential mechanisms for controlling magnetic textures \cite{Wang2024,Khitun2025,Flebus2024}. These forces offer a novel approach to manipulating or stabilizing nanoscale magnetic entities without external fields or spin currents, which is particularly advantageous for the development of future spintronic devices. The concept of Casimir-like interactions presents a promising avenue in this regard. However, despite the potential advantages of these interactions, there remains a challenge in identifying clear experimental evidence for the magnonic Casimir effect. This difficulty primarily arises from the complexity of isolating the contributions of fluctuations from the dominant background interactions, which often obscure the subtle effects being studied.

In this Letter, we investigate the magnonic Casimir effect in antiferromagnetic moiré bilayers. Our emphasis is placed on understanding how the geometrical configurations, dispersion relations, and the magnetic texture influence both the magnitude and direction of this interaction. Within the linear spin-wave theory
we determine the energy shifts induced by fluctuations and pinpoint experimentally viable conditions at which this effect becomes notably significant. The findings underscore the complex interplay among geometric factors, interference patterns, and magnonic vacuum fluctuations, thereby positioning the magnonic Casimir effect as a novel and promising avenue of exploration in spin-wave-mediated interactions.

{\it Basic principle.--} Magnonic zero-point fluctuations represent a fundamental characteristic of quantum behavior within spin systems, persisting even when temperatures reach absolute zero. This persistence highlights the inherently quantum-mechanical nature of magnets, in which the ground state is never entirely motionless \cite{Auerbach1998}. In magnets with ordered structures, such fluctuations have their roots in the Heisenberg uncertainty principle, which governs the relationship between conjugate components of spin, resulting in a non-zero population of virtual magnons that modify the classical spin arrangement. The scale of these fluctuations is intricately linked to exchange interactions, anisotropies, and the system's dimensionality, which can play a pivotal role in altering crucial observables such as the sublattice magnetization, excitation spectrum, and effective interaction parameters \cite{Fazekas1999}. In systems characterized by low dimensionality or frustration, the motion of zero-point magnons is particularly pronounced, potentially disrupting long-range order or leading to the emergence of new quantum phases. The significance of magnon zero-point fluctuations extends beyond theoretical interest; they form a foundational vacuum state from which various phenomena, such as Casimir-like torques \cite{RoldanMolina2015, Tapia2024}, quantum spin pumping \cite{Nunez2014, Ulloa2017, Castro2026}, and hybrid magnon-photon ground-state interactions \cite{Lee2023Topological, Le2025}, arise, thereby cementing their importance in the field of contemporary quantum magnonics \cite{Yuan2022}. 
\begin{figure*}[htb]
\centering
\includegraphics[width=0.99\linewidth]{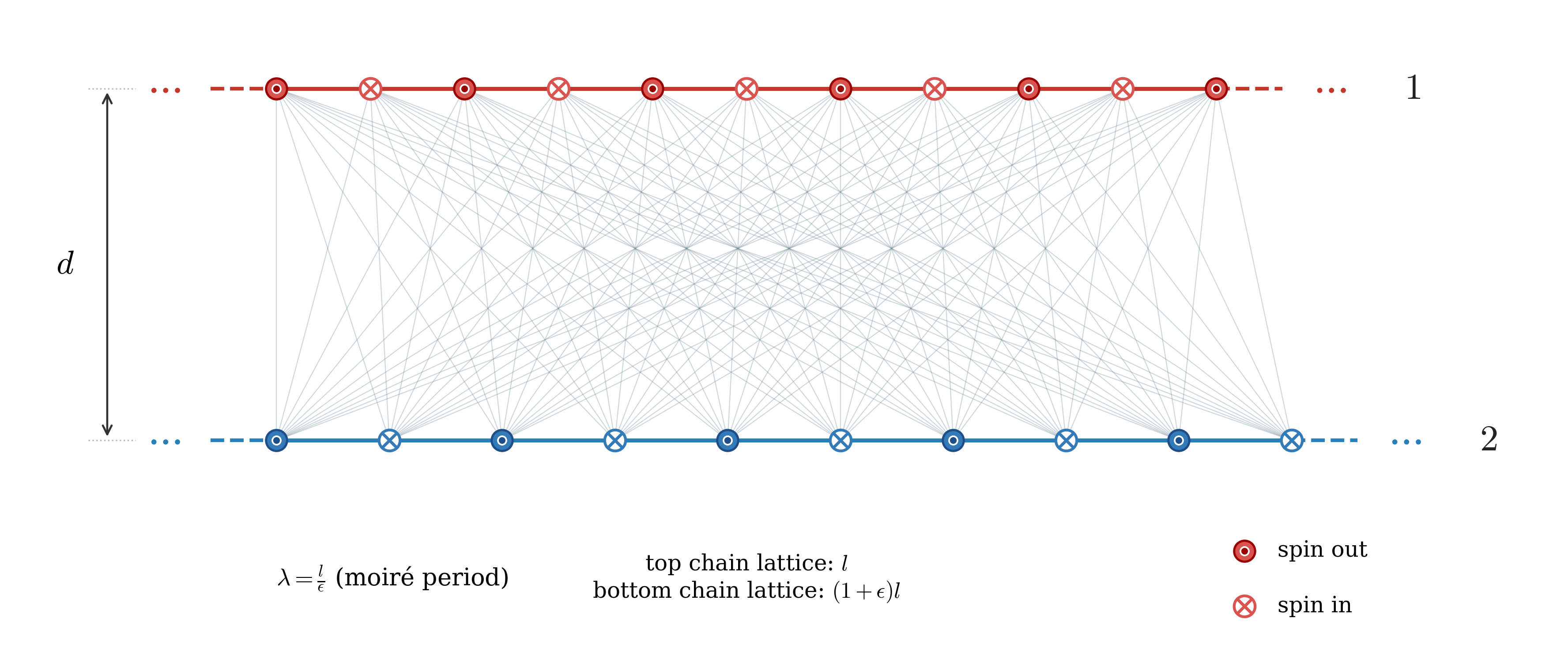}
\caption{Antiferromagnetic moiré ladder consisting of two coupled spin chains, each with antiferromagnetic intrachain exchange $J_0>0$, but with slightly incommensurate lattice constants $a$ and $\ell'=\ell/(1+\varepsilon)$. The mismatch
$\varepsilon$ generates a slowly varying spatial modulation of the interchain (rung) coupling $J_\perp$ at the moiré wavelength $\lambda_M = \ell/\varepsilon$, since the relative alignment between facing sites drifts quasi-periodically
along the ladder.}
    \label{fig:moire}
\end{figure*}

{\it Microscopic model.--} We consider two one-dimensional lattices whose sites $\{\mathbf{R}^1_\mathbf{n}\}$ and $\{\mathbf{R}^2_\mathbf{n}\}$ are decorated by localized magnetic moments $\mathbf{S}_{\mathbf{R}^1_\mathbf{n}}$ and $\mathbf{S}_{\mathbf{R}^2_\mathbf{n}}$. The spin Hamiltonian, $\mathcal{H}=\mathcal{H}_E+\mathcal{H}_A$, contains the contribution of Heisenberg exchange and local easy-axis anisotropy \cite{Coey2012, Auerbach1998}, respectively. On the one hand 
$
\mathcal{H}_E=\mathcal{H}_1+\mathcal{H}_2+\mathcal{H}_{\rm int},
$
 where $\mathcal{H}_1=\sum_{\mathbf{n,m}} J(\mathbf{R}^1_\mathbf{n}-\mathbf{R}^1_\mathbf{m})\mathbf{S}_{\mathbf{R}^1_\mathbf{n}}\cdot\mathbf{S}_{\mathbf{R}^1_\mathbf{m}}$, is the energy contribution associated to the first lattice, $\mathcal{H}_2=\sum_{\mathbf{n,m}} J^\prime(\mathbf{R}^2_\mathbf{n}-\mathbf{R}^2_\mathbf{m})\mathbf{S}_{\mathbf{R}^2_\mathbf{n}}\cdot\mathbf{S}_{\mathbf{R}^2_\mathbf{m}}$, corresponds to the contribution of the second lattice,  and, finally,  $\mathcal{H}_{\rm int}=\sum_{\mathbf{n,m}} \bar{J}(\mathbf{R}^1_\mathbf{n}-\mathbf{R}^2_\mathbf{m})\mathbf{S}_{\mathbf{R}^1_\mathbf{n}}\cdot\mathbf{S}_{\mathbf{R}^2_\mathbf{m}}$ stands for the interaction between the subsystems. In the latter expressions, $J(R)$, $J^\prime(R)$, and $\bar{J}(R)$ are rapidly decaying functions of the distance. For definiteness, we will assume $J(R)=J^\prime(R)=J_0 \exp\beta\left(1-{R}/{\ell}\right)$  and $\bar{J}(R)=J_\perp \exp\beta\left(1-{R}/{\ell}\right)$ where $\ell$ is the characteristic nearest-neighbor distance. We will assume $J_0= J_\perp>0$, leading to an antiferromagnetic ground state.
Additionally, the anisotropy Hamiltonian amounts to adding ${\cal H}_A=-{K}\sum_{\bf n} (\bS^z_{\bf R_n})^2/2$, where $K$ is a constant anisotropy strength.

In the antiferromagnetic system, the Holstein--Primakoff (HP) \cite{Holstein1940} expansion is applied after accounting for the two-sublattice structure of the N\'eel state. For reference, they are usually labeled $A$ and $B$. Because neighboring spins point in opposite directions, one first performs a local rotation on one of the sublattices (typically sublattice $B$) so that all spins become aligned along a common quantization axis \cite{Auerbach1998}. After this rotation, which maps the antiferromagnetic order into an effectively ferromagnetic configuration, the HP transformation is introduced on each site,
$
S_i^{A+} = \sqrt{2S - a_i^\dagger a_i}\, a_i, \qquad
S_i^{A-} = a_i^\dagger \sqrt{2S - a_i^\dagger a_i}, \qquad
S_i^{Az} = S - a_i^\dagger a_i,
$
with analogous expressions for the rotated spins of sublattice B $
S_i^{B-} = \sqrt{2S - b_i^\dagger b_i}\, b_i, \qquad
S_i^{B+} = b_i^\dagger \sqrt{2S - b_i^\dagger b_i}, \qquad
S_i^{Bz} = -S + b_i^\dagger b_i.
$
Expanding the square roots in powers of \(1/S\) and keeping only the quadratic terms yields linear spin-wave theory, where the two sublattices give rise to two bosonic modes that mix through the exchange coupling. In the absence of anisotropy, diagonalizing this quadratic Hamiltonian via a Bogoliubov transformation \cite{Colpa1978} reveals gapless Goldstone modes associated with the spontaneous breaking of isotropy. Higher-order terms in the HP expansion describe magnon--magnon interactions and provide systematic quantum corrections to the order parameter and excitation spectrum of the N\'eel state.

{\it Moiré bands.--} A one-dimensional (1D) moiré pattern emerges when two periodic structures with different lattice constants are superimposed along the same spatial direction. 
This long-wavelength modulation acts as an emergent superlattice, strongly influencing wave propagation, electronic states, and collective excitations in 1D systems.   The system is achieved in our model by setting ${\bf R}^{1}_n=n\,\ell\,\boldsymbol{x}$ and  ${\bf R}^2_n=n\,(1+\varepsilon)\,\ell\,\boldsymbol{x}+d\boldsymbol{y}$, as in Ref. \cite{Patra2026, Wang2021, Mellado2026}, while the sublattices $A$ and $B$ are related to the parity of the index $n$. The periodicity of the emergent moiré pattern is $\Lambda_M=(1+\varepsilon)\,\ell/\varepsilon$. A schematic drawing of this construction is rendered in Fig.~\ref{fig:moire}.

\begin{figure}
    \centering
    \includegraphics[width=1.0\linewidth]{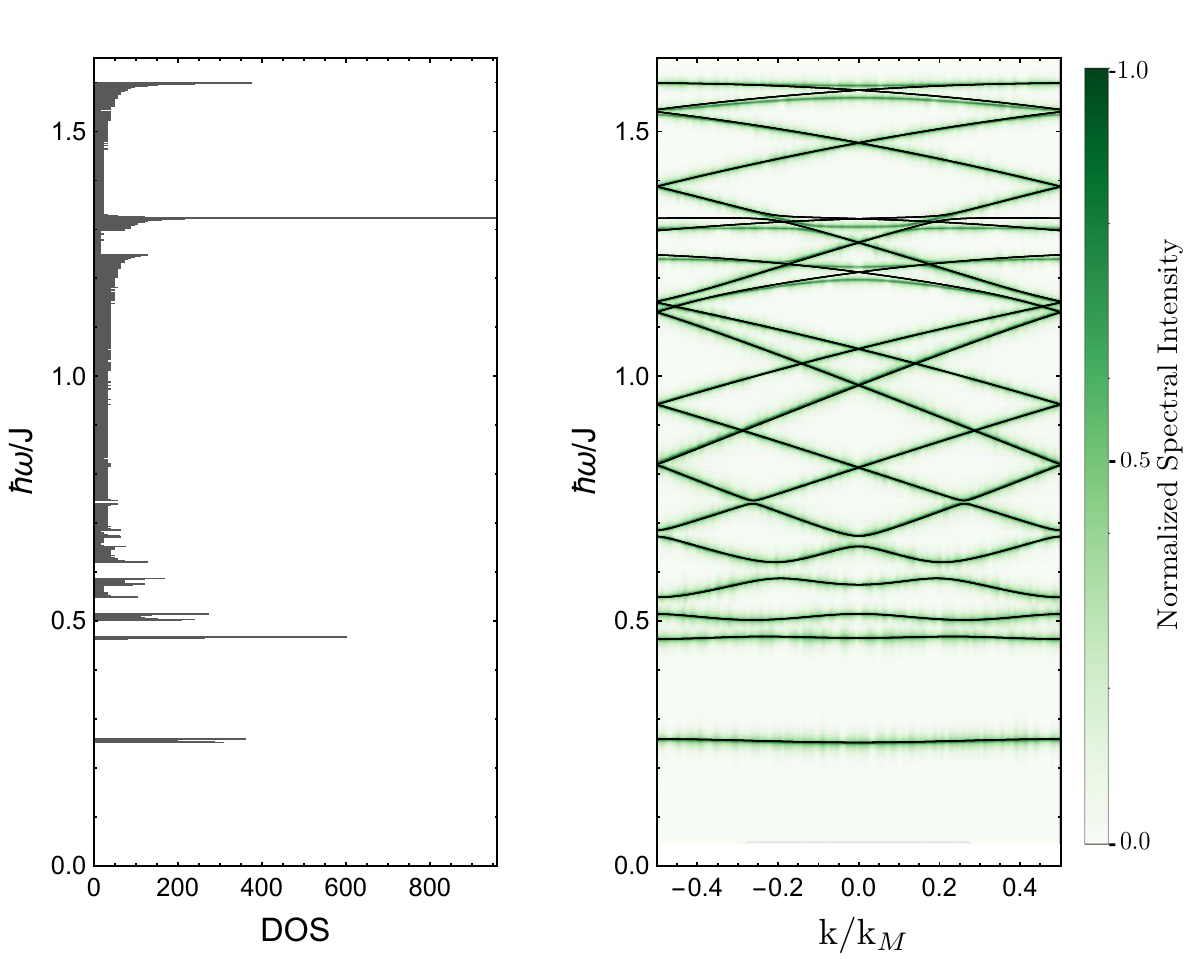}
    \caption{Right: Magnonic bands within the moiré Brillouin zone for the case $\epsilon=0.1$, $J_0=J=$ unit of energy, $J_\perp = J $, $\beta=1$, $K= 0.025\; J$. The moiré Brillouin zone lies within $[-G_M/2,G_M/2]$, where $G_M=2\pi/\Lambda_M$. The band structure displays the characteristic moiré splitting. In the green background, we display the spectral function obtained using Landau-Lifshitz dynamics.  Left: The density of states is consistent with the band structure. Many bands overlap and cross, and show pronounced peaks---van Hove singularities---at the energies of the flat-band extrema.}
    \label{fig: bandas}
\end{figure}

We follow the seminal work of Bistritzer-MacDonald (BM) \cite{Bistritzer2011} to address the problem. A stark difference between the present problem and the one in twisted graphene is that we are currently dealing with bosonic excitations. All bands are relevant, not only those near the Fermi level. This makes it critical to determine how many moiré replicas to use. After Bogoliubov diagonalization, the BM Hamiltonian can be written as:
$$
\mathcal{H}=\sum_{k,n}\hbar\omega^{\phantom\dagger}_{nk}\left(\boldsymbol{\alpha}^\dagger_{nk}\boldsymbol{\alpha}^{\phantom\dagger}_{nk}+g^{\phantom\dagger}_{nk}\right),
$$
where $\boldsymbol{\alpha}$'s fulfill a bosonic algebra, $k$ lies over the Moiré Brillouin zone, and $n$ is the band index.  The vacuum state is defined as being annihilated by the destruction operators, namely $\boldsymbol{\alpha}^{\phantom\dagger}_{nk}|\textrm{vac}\rangle=0$ for all $k$ and $n$. Just like in other bosonic problems, the $g^{\phantom\dagger}_{nk}$ are some constants quantifying the zero-point energy contribution from each mode, which depart from the standard ${1}/{2}$ that is canonical in the harmonic oscillator problem. The energy bands $\mathcal{E}_{nk}=\hbar\,\omega_{nk}$ over the moiré Brillouin zone are displayed in Fig. \ref{fig: bandas}. We compare them with results from classical Landau-Lifshitz (LL) dynamics. The agreement is remarkable. The magnonic band structures obtained from the LL framework and HP approach exhibit broad agreement but differ in several ways. The LL equation describes classical spin dynamics, retaining the full nonlinearity of precession, whereas the HP transformation maps spin operators onto bosonic magnon operators and is truncated at quadratic order, corresponding to linear spin-wave theory. This truncation neglects magnon-magnon interactions, which can renormalize band energies and modify gap openings at points where modes hybridize. Consequently, avoided crossings and symmetry-protected gaps that arise from nonlinear dynamical effects may be captured by LL but absent in the linearized HP spectrum, or vice versa when gaps owe their origin to quantum zero-point fluctuations, which have no classical analog and therefore do not appear in LL. 

\begin{figure}
    \centering
    \includegraphics[width=1.0\linewidth]{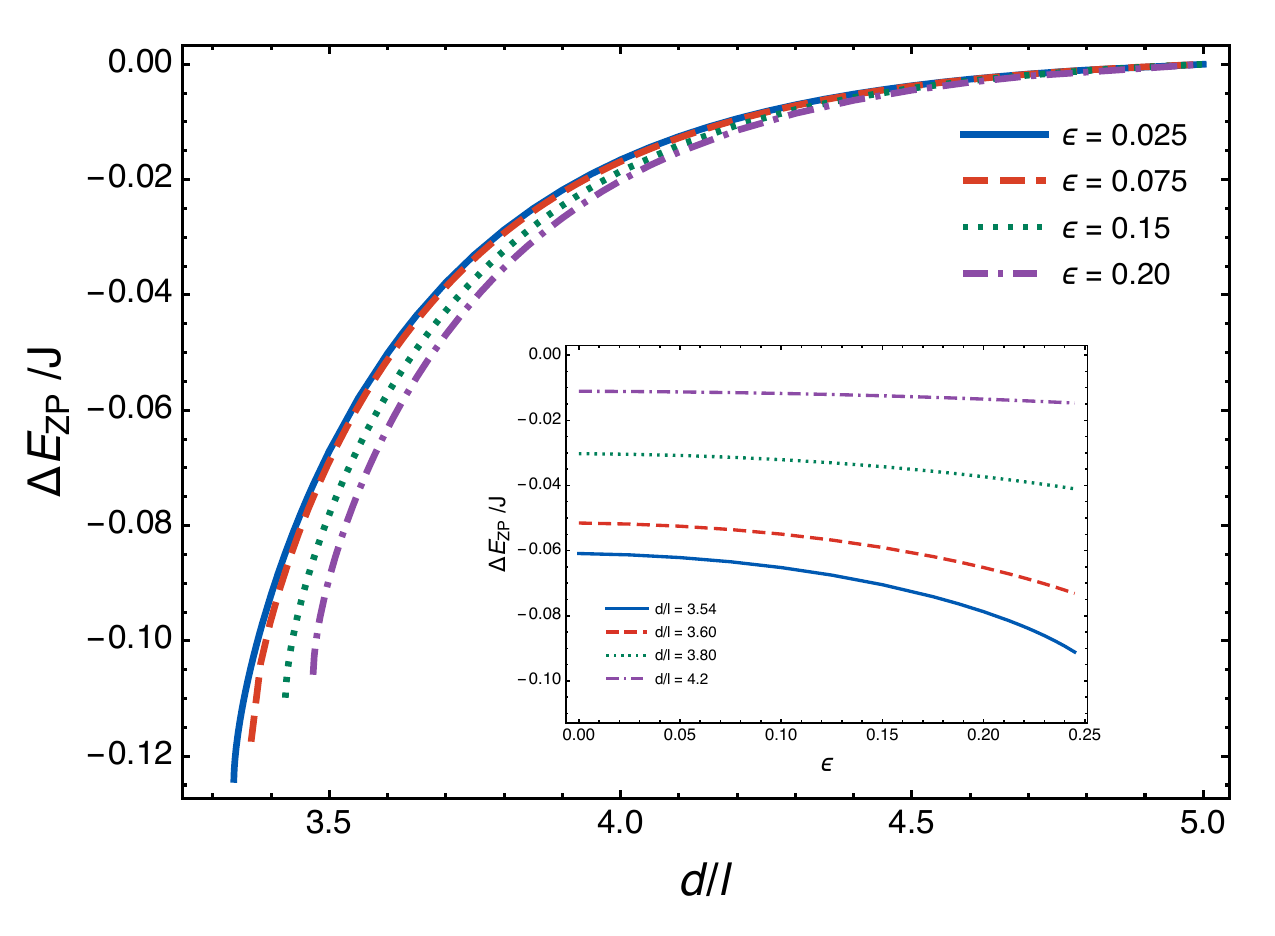}
    \caption{Zero-point energy, $\Delta E_{ZP}=E_{ZP}(d,\varepsilon)-E_{ZP}(\infty,\varepsilon)$, versus $d/\ell$ for various values of $\varepsilon$. Inset: $E_{ZP}(d,\varepsilon)$ as a function of $\varepsilon$ for various values of $d$. }
    \label{fig: Ezp}
\end{figure}

{\it Moiré-based magnonic zero-point energy.--} The Casimir effect, originally predicted for the electromagnetic vacuum between conducting plates \cite{Casimir1948}, has been generalized to a broad class of bosonic fields whose zero-point fluctuations are modified by boundary conditions. In magnetic insulators, spin-waves (magnons) constitute a bosonic field whose vacuum fluctuations can give rise to an analogous {magnonic Casimir effect} \cite{Nakata2023,Tapia2024}. When two interfaces that impose boundary conditions on the magnon field---such as surfaces of a ferromagnetic slab, domain walls, or the edges of a magnonic waveguide---are brought into close proximity, the discrete magnon spectrum between them differs from the continuum of modes outside, generating a net force. Unlike its electromagnetic counterpart, the magnonic Casimir force is highly tunable: the magnon dispersion can be controlled by an external magnetic field, uniaxial anisotropy, or exchange stiffness, offering additional knobs absent in photonic systems. Furthermore, the thermal occupation of magnon modes at finite temperature introduces a substantial classical (Lifshitz-like) contribution even at room temperature \cite{Lifshitz1956, Dzyaloshinskii1961}, making magnonic Casimir forces experimentally more accessible than their electromagnetic analogs at the nanoscale. The vacuum expectation value of the energy is $E_{ZP}=\langle \textrm{vac}|\mathcal{H}|\textrm{vac}\rangle=\sum_{n k} \hbar\, \omega^{\phantom\dagger}_{nk}g^{\phantom\dagger}_{nk}$. These values are calculated and displayed for different geometrical parameters $d$ and $\varepsilon$ and displayed in Fig. \ref{fig: Ezp}. Such dependency, $E_{ZP}(d,\varepsilon)$, constitutes the main result of this work.

Recent theoretical work has placed the magnonic Casimir effect on a firmer microscopic footing by deriving the fluctuation-induced pressure directly from the spin-wave Hamiltonian using functional integral and scattering-matrix methods~\cite{Nakata2023}. In quasi-one-dimensional magnonic structures---including spin chains and helimagnetic wires---the interplay between exchange interactions, dipolar fields, and geometric confinement leads to a rich dependence of the Casimir force on the separation $d$ between boundaries, interpolating between $d^{-3}$ and $d^{-4}$ power laws depending on the dominant retardation regime and dimensionality. In our case, fitting to our data provides the long-distance law $E_{ZP}(d,\varepsilon)\sim E_{ZP}(\infty,\varepsilon)+f(\varepsilon)\;d^{-\alpha}$, where $\alpha \sim 9$ independent of the value of $\varepsilon$. The possibility of sign reversal of the force---from attractive to repulsive---by tuning the magnetic anisotropy or by introducing chiral interactions (Dzyaloshinskii--Moriya interaction) has been highlighted as a distinctive feature with no direct photonic analogue. Beyond the scalar pressure, fluctuation-induced angular momentum transfer between magnetic bodies gives rise to a {Casimir spin torque}: vacuum and thermal magnon fluctuations carry spin angular momentum that is exchanged between non-collinear or asymmetric magnetic slabs, exerting a torque that tends to reorient their magnetizations even in the absence of any direct exchange coupling~\cite{RoldanMolina2015, Tapia2024}. This torque is the reciprocal of the Casimir force in the sense that both originate from the same fluctuation spectrum, yet it is an independent observable that couples to magnetization dynamics and could, in principle, be detected via ferromagnetic resonance frequency shifts or magnon-mediated switching in multilayer geometries.

\begin{figure}
    \centering
    \includegraphics[width=1.0\linewidth]{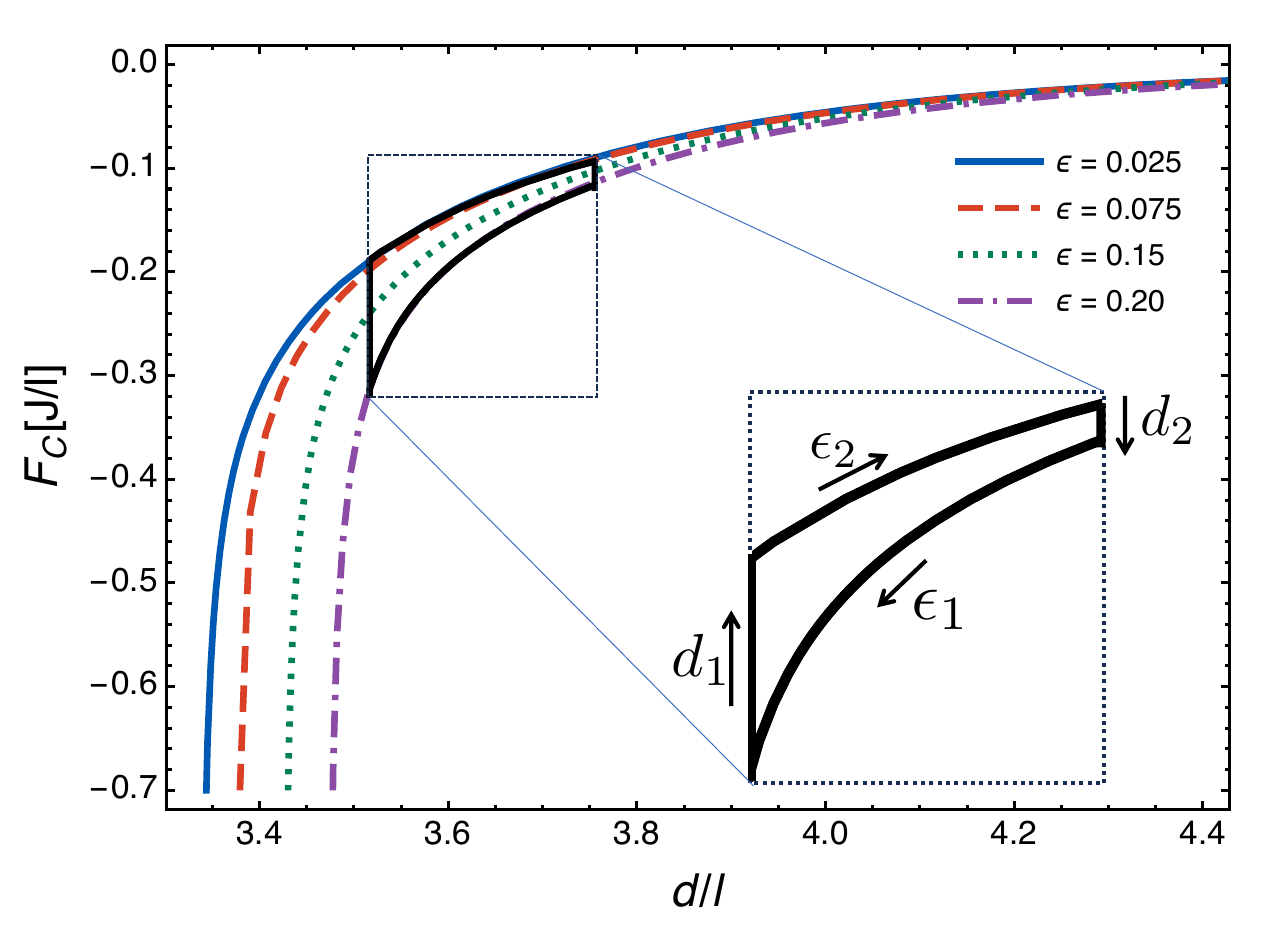}
    \includegraphics[width=1.0\linewidth]{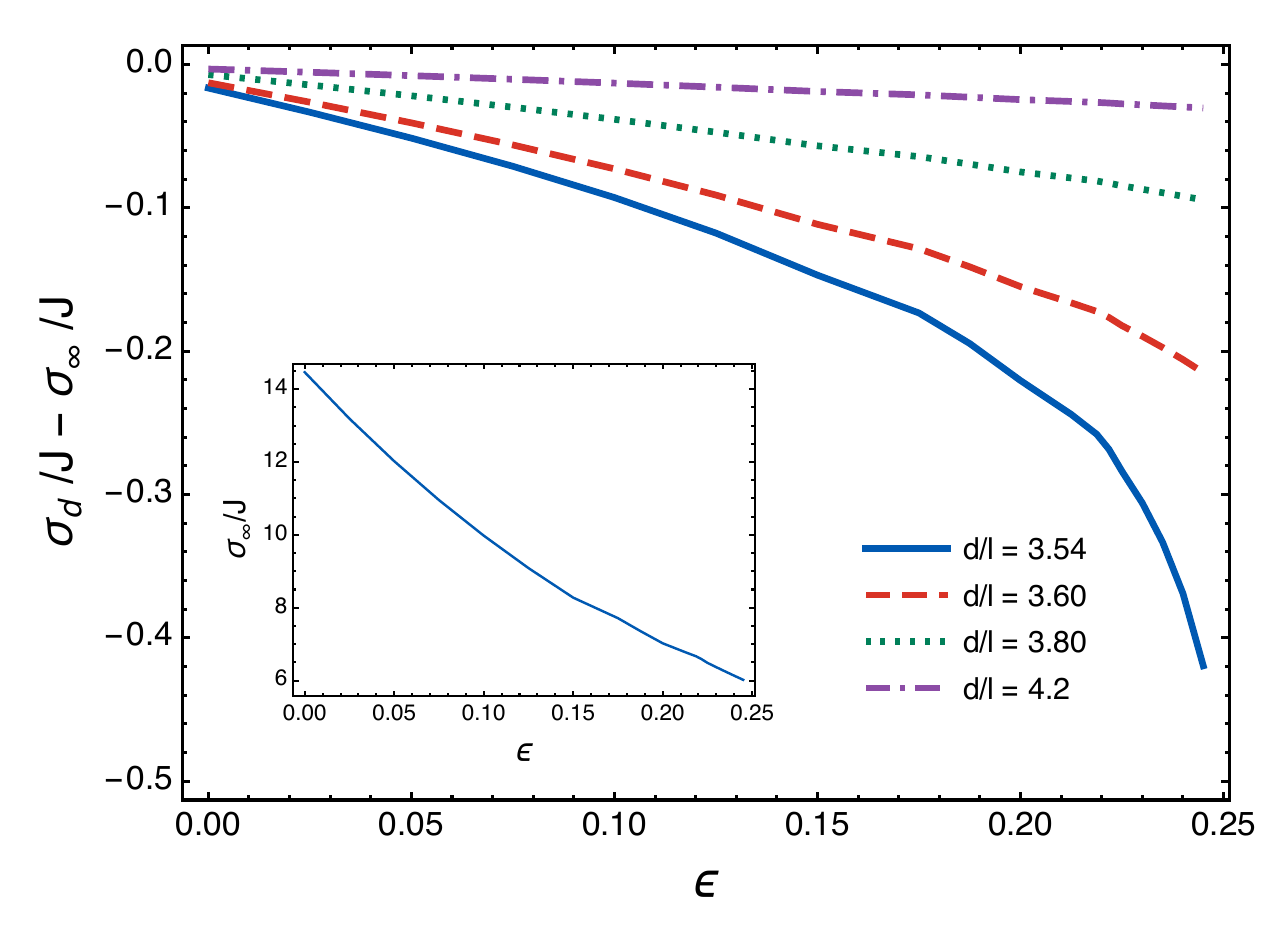}
    \caption{Top: Casimir's force, $F_C$, for various values of $\varepsilon$. Bottom: Casimir’s stress, $\sigma_C(d,\varepsilon)-\sigma_C(\infty,\varepsilon)$, for various values of $d$. Inset: $\sigma_C(\infty,\varepsilon)$ vs $\varepsilon$.  }
    \label{fig: Force and stress}
\end{figure}

{\it Casimir force and stress, and mechanical cycle.--} The zero-point energy is defined as a function of $d$ the inter-chain space and $\varepsilon$ the relative strain between the chains, $E_{ZP}(d,\varepsilon)$. In this way, in an analogous fashion to the Casimir's force, $F_C=-\left.({\partial E_{ZP}}/{\partial d})\right|_{\varepsilon}$, we can define the Casimir's stress $\sigma_C=-\left.({\partial E_{ZP}}/{\partial \varepsilon})\right|_{d}$. This contribution reflects the system's reluctance, due to vacuum fluctuations, to undergo relative changes in the ladder's elastic deformation. The behavior of the Casimir's forces and stresses as a function of $d$ and $\varepsilon$ is depicted in Fig. \ref{fig: Force and stress}. Despite our results spanning a small range of distances, we can identify two regimes. In the long-distance regime, the behavior follows a power law, such as $F_C\sim -\alpha f(\varepsilon)/d^{1+\alpha}$. However, in the short-distance regime, the behavior is modified by a critical increment in the form $F_C\sim f_0 |d-d_c(\varepsilon)|^{-\gamma}$, where our estimates indicate that $\gamma\sim 0.5$. For values of $d$ closer than $d_c(\varepsilon)$, our methodology fails because the system is not definite positive. 

We propose a mechanical cycle that runs on the principle of converting zero-point magnetic energy into elastic energy. Starting from the configuration point $(d_1,\varepsilon_1)$, we move the chains closer together, keeping the strain constant, toward configuration point $(d_2,\varepsilon_1)$. Next, we stretch the ladder, at constant separation, and move the system to the point $(d_2,\varepsilon_2)$. The next stage naturally consists of separating the ladder to the point $(d_1,\varepsilon_2)$. Finally, we close the cycle by expanding the ladder while keeping the separation constant. During the separation and approach,  work is done against and in favor of the Casimir's force, respectively. The net value is:
$$ W=W_1-W_2=\int^{d_2}_{d_1}(F_C(x,\varepsilon_1)-F_C(x,\varepsilon_2)){\rm d}x,$$
Meanwhile, during expansion and compression, work is done against and in favor of the Casimir's stress, respectively. The net value is:
$$
Q=Q_1-Q_2=\int^{\varepsilon_2}_{\varepsilon_1}(\sigma_C(d_1,\varepsilon)-\sigma_C(d_2,\varepsilon)){\rm d}\varepsilon,
$$ 
As can be seen in Fig. \ref{fig: Force and stress}. Naturally, we have $Q+W=0$, as it is the case for any loop integral of an exact differential.  We define the nominal efficiency of the cycle as $\eta=W/Q_2=1-Q_1/Q_2$ \cite{Fermi1956, Callen1985}; by taking $d_2\gg \ell/\beta$, we can make this efficiency approach 1, signaling an optimal energy transducer.

This closed loop in displacement–strain phase space could serve as the working principle for a new type of mechanical actuator, energy harvester, or even a mechanical engine. The key idea is that a cyclic trajectory in the two-dimensional space can produce a net transfer of energy, even though both degrees of freedom return to their initial values after each cycle.  This "motor" can be turned into a useful vacuum-based magneto-electrical device by coupling the strain channel to piezoelectric \cite{Heywang2008, Yang2019, Willatzen2024} or piezospintronic \cite{Nunez2014, Ulloa2017, Saez2024, Chen2025, Castro2026} signals.

{\it Discussion and conclusions.--} 
In this work, we have presented a comprehensive analysis of the magnonic Casimir effect in antiferromagnetic moir\'e bilayers, combining a microscopic HP treatment of the N\'eel state with the emergent band structure imposed by the incommensurate one-dimensional superlattice. The central finding is that the moir\'e geometry acts as a powerful and tunable modulator of the magnonic vacuum: by controlling the lattice mismatch parameter $\varepsilon$, one continuously reshapes the magnon spectrum—its sub-band widths, the density and magnitude of moir\'e-induced gaps, and the location of flat-band-like regions near the moir\'e Brillouin zone edges—thereby engineering the zero-point fluctuation energy and, consequently, the Casimir interaction between the two magnetic layers.

The  gap structure translates directly into a dependence of the Casimir energy on interlayer separation $d$ and  mismatch $\varepsilon$. In particular, the flat-band regions, where the magnon group velocity vanishes, and the density of states is strongly enhanced, give rise to parametrically large contributions to the zero-point energy. This amplification is absent in the ferromagnetic case and constitutes a distinctive signature of the interplay between antiferromagnetic order and moir\'e geometry. 

From a broader perspective, the present results establish magnetic moir\'e systems as a versatile platform for exploring zero-point fluctuation-induced phenomena in quantum magnets. The tunability of the moir\'e period via mechanical strain or twist angle~\cite{Patra2026, Mellado2026}, combined with the intrinsic robustness of antiferromagnetic order against external magnetic
perturbations, makes these systems particularly well suited for experimental investigation. Candidate material realizations include van der Waals antiferromagnets such as \ce{MnPS_3}, \ce{FePS_3}, or \ce{CrI_3} bilayers assembled with a controlled lattice mismatch, where recent advances in deterministic stacking techniques have demonstrated the feasibility of engineering moir\'e superlattices with atomic-scale precision.
Generalizations of our results and proposals to other moiré patterns and even other bosonic excitations are straightforward. They will be the matter of future explorations from our group.


{\it Acknowledgments.--} This work was funded by ANID CEDENNA CIA 250002.  Funding is acknowledged from Fondecyt Regular 1230515 and 1230747.

\bibliography{casimir_moire}
\end{document}